\documentclass{emulateapj}
\usepackage{apjfonts}

\usepackage{psfig}
\usepackage{amsmath} 

\submitted{Accepted by the Astrophysical Journal} 
 
\shorttitle{\thepulsar: An Eclipsing Binary Radio MSP}

\shortauthors{Crawford et al.} 
 
\newcommand{\thepulsar}{PSR~J1723$-$2837}
\newcommand{\rosatsource}{1RXS~J172323.7$-$283805}
\newcommand{\integralsource}{IGR~J17233$-$2837}

\begin{document}
 
\title{\thepulsar: An Eclipsing Binary Radio Millisecond Pulsar}

\author{Fronefield Crawford\altaffilmark{1}, 
Andrew G. Lyne\altaffilmark{2},
Ingrid H. Stairs\altaffilmark{3},
David L. Kaplan\altaffilmark{4,5},
Maura A. McLaughlin\altaffilmark{6},
Paulo C. C. Freire\altaffilmark{7},
Marta Burgay\altaffilmark{8},
Fernando Camilo\altaffilmark{9,10},
Nichi D'Amico\altaffilmark{8},
Andrew Faulkner\altaffilmark{11},
Michael Kramer\altaffilmark{7},
Duncan R. Lorimer\altaffilmark{6,12},
Richard N. Manchester\altaffilmark{13}, 
Andrea Possenti\altaffilmark{8},
Danny Steeghs\altaffilmark{14}}

\altaffiltext{1}{Department of Physics and Astronomy, Franklin and
Marshall College, P.O. Box 3003, Lancaster, PA 17604, USA; email:
fcrawfor@fandm.edu}
 
\altaffiltext{2}{Jodrell Bank Centre for Astrophysics, University of
Manchester, Manchester M13 9PL, UK}

\altaffiltext{3}{Department of Physics and Astronomy, University of
British Columbia, 6224 Agricultural Road, Vancouver, BC V6T 1Z1,
Canada}

\altaffiltext{4}{Physics Department, University of Wisconsin -
Milwaukee, Milwaukee, WI 53211, USA}

\altaffiltext{5}{Department of Astronomy, University of Wisconsin -
Madison, Madison, WI 53715, USA}

\altaffiltext{6}{Department of Physics, West Virginia University,
Morgantown, WV 26506, USA}

\altaffiltext{7}{Max-Planck-Institut f\"{u}r Radioastronomie, auf dem
Huegel 69, 53121 Bonn, Germany}

\altaffiltext{8}{INAF - Osservatorio Astronomico di Cagliari, Poggio
dei Pini, 09012 Capoterra, Italy}

\altaffiltext{9}{Columbia Astrophysics Laboratory, Columbia
University, New York, NY 10027, USA}

\altaffiltext{10}{Arecibo Observatory, HC3 Box 53995, Arecibo, PR
00612, USA}

\altaffiltext{11}{Cavendish Laboratory, University of Cambridge, J. J.
Thompson Avenue, Cambridge, CB3 0HE, UK}

\altaffiltext{12}{National Radio Astronomy Observatory, P.O. Box 2,
Green Bank, WV 24944, USA}

\altaffiltext{13}{CSIRO Astronomy and Space Science, Australia
Telescope National Facility, P.O. Box 76, Epping, NSW 1710, Australia}

\altaffiltext{14}{Department of Physics, University of Warwick,
Coventry CV4 7AL UK}

\begin{abstract}
We present a study of \thepulsar, an eclipsing, 1.86 ms millisecond
binary radio pulsar discovered in the Parkes Multibeam survey.  Radio
timing indicates that the pulsar has a circular orbit with a 15 hr
orbital period, a low-mass companion, and a measurable orbital period
derivative.  The eclipse fraction of $\sim 15$\% during the pulsar's
orbit is twice the Roche lobe size inferred for the companion.  The
timing behavior is significantly affected by unmodeled systematics of
astrophysical origin, and higher-order orbital period derivatives are
needed in the timing solution to account for these variations.  We
have identified the pulsar's (non-degenerate) companion using archival
ultraviolet, optical, and infrared survey data and new optical
photometry.  Doppler shifts from optical spectroscopy confirm the
star's association with the pulsar and indicate a pulsar-to-companion
mass ratio of $3.3 \pm 0.5$, corresponding to a companion mass range
of 0.4 to 0.7 $M_{\odot}$ and an orbital inclination angle range of
between 30 and $41^{\circ}$, assuming a pulsar mass range of 1.4$-$2.0
$M_{\odot}$. Spectroscopy indicates a spectral type of G for the
companion and an inferred Roche-lobe-filling distance that is
consistent with the distance estimated from radio dispersion.  The
features of \thepulsar\ indicate that it is likely a ``redback''
system.  Unlike the five other Galactic redbacks discovered to date,
\thepulsar\ has not been detected as a $\gamma$-ray source with {\it
Fermi}. This may be due to an intrinsic spin-down luminosity that is
much smaller than the measured value if the unmeasured contribution
from proper motion is large.
\end{abstract}


\section{Introduction}

In the standard model for millisecond pulsar (MSP) production
\citep{acr+82}, old neutron stars are recycled by the accretion of
matter from a companion star. This model predicts that the longer this
phase of the evolution lasts, the shorter the period of the resulting
MSP will be.  During the accretion phase, copious X-ray emission is
produced, and the neutron star is observed as a low-mass X-ray binary
(LMXB) in cases where the mass of the companion is small. After the
accretion phase is finished and the pulsar has been spun up, the
neutron star may emit detectable radio pulses and be observable as an
MSP with an evolved white dwarf companion. This evolutionary path from
LMXB to radio MSP is important to verify and understand. The discovery
of the first accretion-powered MSP in X-rays, SAX 1808.4$-$3658
\citep{cm98}, suggested this connection. However, objects that are
actually undergoing a transition between a LMXB phase and a radio MSP
phase further solidify this link, but these systems are rare.  It was
not until the discovery of PSR J1023+0038 by \citet{asr+09} that a
transition object was discovered.  Since then, a total of five objects
(called ``redbacks'') having similar companion characteristics and
eclipsing behavior have been discovered in our Galaxy \citep{r13}. In
2013, a redback system in the globular cluster (GC) M28, PSR~J1824$-$2452I,
was found to swing between being a radio MSP and an LMXB with detectable
X-ray pulsations at the exact period and orbital characteristics of
the radio pulsar (an accretion-powered MSP). It then went back to being a 
radio MSP,
confirming the evolutionary link beyond doubt and establishing the
existence of an intermediate phase where the accretion and radio MSP
phases alternate on very short timescales \citep{pfb+13}.

\thepulsar\ is a radio MSP that was discovered by \citet{fsk+04} in
the Parkes multibeam pulsar survey.  It has a spin period of 1.86 ms,
making it the 12th most rapidly rotating radio pulsar currently listed
in the ATNF Pulsar Catalogue
\citep{mht+05}\footnote{http://www.atnf.csiro.au/research/pulsar/psrcat/,
V1.46}.  From the original survey detection, it was clear that the
pulsar was highly accelerated and flux-variable, making it difficult
to detect despite being very bright at certain orbital phases.  In
this paper we present and discuss the timing results for \thepulsar\
and optical observations that identify and characterize the pulsar's
companion.  \thepulsar\ is the sixth Galactic redback yet discovered,
and this number will most likely grow as radio surveys continue to
detect new pulsar systems identified in {\it Fermi} $\gamma$-ray
observations \citep{rap+12}.

The paper is laid out as follows. In Section 2, we describe the timing
observations of \thepulsar, and in Section 3 we present the timing
results.  In Section 4, we describe archival and new optical and
infrared observations that identify the companion star and demonstrate
its association with the pulsar.  Section 5 presents a discussion of
the system, and Section 6 presents our conclusions and summarizes our
results.

\section{Timing Observations}

Between March 2001, when \thepulsar\ was discovered, and September
2004, the pulsar was only sporadically detected in 1400 MHz
observations that were attempted with the Parkes 64-m telescope.  In
September 2009, we made several new detections with exploratory
observations at 2000 MHz using the Robert C. Byrd Green Bank Telescope
(GBT). We derived a preliminary orbital ephemeris from these Parkes
and GBT detections using a method described by \citet{fkl01} which
uses the measured spin periods and accelerations of the pulsar and is
well-suited for cases where detections are sparse.  Using this new
orbital solution, timing observations resumed in December 2009 with
the Jodrell Bank 76-m Lovell telescope and in February 2010 with the
GBT.  Between September 2009 and March 2011, the pulsar was detected
regularly at frequencies of 1520 MHz (with Jodrell), 2000 MHz (with
the GBT), and also several times at 1369 and 3100 MHz (with Parkes).
These observations were used to obtain a phase-connected timing
solution, which we present here. For all of the observations
discussed, dual polarizations were summed at the telescope prior to
data recording and analysis.

\begin{deluxetable*}{lcccccc}
\tablecaption{Timing Observations of \thepulsar\label{tbl-1}}
\tablewidth{0pt}
\tablehead{
\colhead{Telescope} &
\colhead{Backend} &
\colhead{Center Freq.} &
\colhead{$N_{\rm TOA}$\tablenotemark{a}} &
\colhead{MJD} &
\colhead{TEMPO\tablenotemark{b}} &
\colhead{Detection\tablenotemark{c}}\\
\colhead{} &
\colhead{} &
\colhead{(MHz)} &
\colhead{} &
\colhead{Range} &
\colhead{EFAC} &
\colhead{Percentage}
}
\startdata
GBT          & GUPPI & 2000 & 234 & 55102$-$55263 & 2.15 & 75\%    \\
             &       &      &     &               &      &         \\
Jodrell Bank & DFB   & 1520 & 138 & 55196$-$55803 & 14   & 30\%    \\
             &       &      &     &               &      &         \\ 
Parkes       & PDFB4 & 1369 &   2 & 55292         & 1    & 50\%    \\ 
             & PDFB4 & 3100 &  10 & 55297$-$55312 & 1    & ~~~80\%
\enddata

\tablenotetext{a}{Number of TOAs used in  the timing fit.
Multiple TOAs were generated from each GBT and Parkes
observation.}

\tablenotetext{b}{TOA uncertainties from each instrumental setup were
multiplied by this factor to correct for the generally underestimated
uncertainties produced by the program that generates the TOAs.}

\tablenotetext{c}{Approximate percentage of timing observations in
which the pulsar was detected.}

\end{deluxetable*}
				   	     

Observations with the GBT used the Green Bank Ultimate Processing
Instrument (GUPPI) to observe the pulsar at 2000 MHz in a number of
sessions.  In these observations, a bandwidth of 800 MHz was split
into 2048 filterbank channels, each of which was sampled at 64
$\mu$s. Integration times ranged from about 30 minutes to 2
hr. Multiple times-of-arrival (TOAs) were derived from each
observation using the PRESTO software package \citep{r01,
rem02}. Sub-integrations from each observation were dedispersed and
folded, and the resulting profiles were compared to a pulse profile
template produced from one of the strong detections.  \thepulsar\ was
also observed regularly at a center frequency of 1520 MHz with the
Lovell telescope.  These observations used a cryogenically cooled
receiver and a digital filterbank backend with 384 MHz of bandwidth
split into 1536 frequency channels of width 0.25 MHz.  The sampling
time was 128 samples per pulse period, or 14.5 $\mu$s. Typical
integration times were between 30 and 40 minutes per observation.
TOAs were produced from these observations, and the first
phase-connected timing solution was made with these data.  In April
2010, Parkes observations were conducted for 7 sessions with the PDFB4
digital backend. Two of these sessions were conducted at a center
frequency of 1369 MHz, and five were at 3100 MHz. One 1369 MHz
detection and four 3100 MHz detections were made from these
observations, and TOAs were produced and included in the timing
solution.  Table \ref{tbl-1} shows the details of the successful
timing observations from the various telescopes.

The original set of detections at 1400 MHz taken at Parkes between
2001 and 2004 could not be phase connected with our timing solution
and are not included here.  Additional timing observations in the
future may be useful for phase connecting these older Parkes data, but
the timing jitter and the systematics present in the TOAs may preclude
this (see discussion below).

\section{Timing Results}

Timing data from the three telescopes spanning a total of 522 days
were phase connected using the TEMPO software
package\footnote{http://tempo.sourceforge.net}. The BTX binary model
(D. Nice, unpublished) was used. This is a modification of the BT
binary model \citep{bt76} that allows for multiple derivatives of the
orbital frequency to be included, as is necessary for \thepulsar. We
used the UTC(NIST) time standard and the JPL DE405 solar system
ephemeris for barycentric corrections.  The TOA uncertainties were
adjusted for each observing setup by a multiplicative factor (EFAC in
TEMPO) to account for underestimated values produced in the generation
of the TOAs. Specifically, TOAs from each observing setup were
separately fit to the timing model such that the chosen multiplicative
factor produced a reduced $\chi^{2}$ close to unity. The combined data
set was then fit with these separate factors in place for each subset.
The TOAs were weighted in the comprehensive fit according to their
uncertainties. Systematics present in the residuals required the
inclusion of orbital frequency derivatives up to the third derivative
to whiten the residuals.

\begin{figure*}
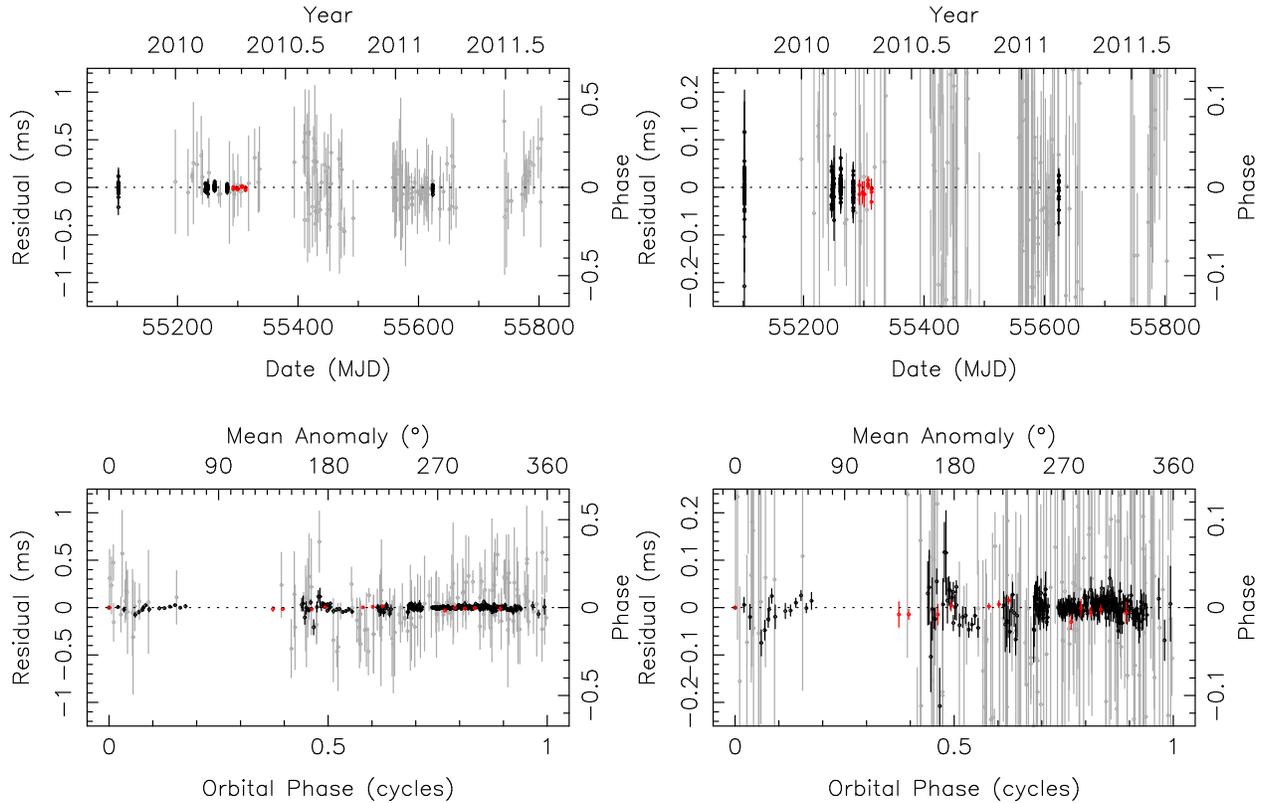

\centerline{\psfig{figure=fig1a.ps,width=3.25in,angle=270} \psfig{figure=fig1c.ps,width=3.25in,angle=270}}
\vspace{0.2in}
\centerline{\psfig{figure=fig1b.ps,width=3.25in,angle=270} \psfig{figure=fig1d.ps,width=3.25in,angle=270}}
\caption{Timing residuals for \thepulsar\ shown as a function of date
(top) and orbital phase (bottom).  The TOAs from the three different
telescopes used are color coded (black for 2000 MHz GBT data; light
gray for 1520 MHz Jodrell Bank data; red for 1369 and 3100 MHz Parkes
data). The plots on the right show the same data as the plots on the left, but
on a smaller scale so that the higher-precision GBT and Parkes TOAs
can more easily be seen.  The timing solution is presented in Table
\ref{tbl-2} and has an rms residual of 21 $\mu$s.  No detections were
made for $\sim 15$\% of the orbit at phases near 0.25, despite some
attempts to observe the pulsar during these phases. This is consistent
with eclipsing by an extended companion envelope.  The TOA
uncertainties for each data subset were scaled to produce a reduced
$\chi^{2}$ close to one for that subset in order to account for the
additional uncertainty introduced by both instrumental and
astrophysical sources. Several higher-order orbital frequency
derivatives were included in the timing solution to whiten the
residuals.\label{fig-1}}
\end{figure*}

The timing solution is presented in Table \ref{tbl-2}, and the quoted
uncertainties in the timing parameters are twice the formal TEMPO
uncertainties. This is common practice when the uncertainties have
been increased by EFAC.  The multiple observing frequencies allowed us
to fit for a DM of 19.688(13) pc cm$^{-3}$. This indicates a
relatively small distance to the pulsar ($d \sim 0.75$ kpc), according
to the NE2001 DM-distance model of \citet{cl02}.  Figure \ref{fig-1}
shows the timing residuals after fitting the TOAs to the timing
model. The residuals are shown as a function of both date and orbital
phase, with a weighted rms of the residuals of 21 $\mu$s.

\begin{deluxetable*}{lc}
\tabletypesize{\scriptsize}
\tablecaption{Timing Parameters for \thepulsar\label{tbl-2}}
\tablewidth{0pt}
\tablehead{
\colhead{Parameter} &
\colhead{Value} 
}
\startdata
Measured Quantities                           &                     \\
\hline \\
Right ascension (J2000)                       &  17:23:23.1856(8)   \\      
Declination (J2000)                           & $-$28:37:57.17(11)  \\ 	         
Spin frequency, $f$ ($s^{-1}$)                & 538.870683485(3)    \\     
Frequency derivative, $\dot{f}$ ($s^{-2}$)    & $-2.19(12) \times 10^{-15}$  \\ 
Dispersion measure, DM (pc cm$^{-3}$)         & 19.688(13)          \\	    	 
Timing epoch (MJD)                            & 55667               \\	     	   	
Time of ascending node, $T_{\rm asc}$ (MJD)   & 55425.320466(2)     \\	      
Projected semi-major axis, $x$ (s)\tablenotemark{a}            & 1.225807(9) \\		       
Orbital frequency, $f_{b}$ ($s^{-1}$)         & $1.88062856(2) \times 10^{-5}$               \\
Orbital frequency derivative, $\dot{f_{b}}$ ($s^{-2}$) & $1.24(4) \times 10^{-18}$          \\
Orbital frequency second derivative, $\ddot{f_{b}}$ ($s^{-3}$) & $-2.6(3) \times 10^{-26}$   \\
Orbital frequency third derivative, $\dddot{f_{b}}$ ($s^{-4}$) & $-6.2(6) \times 10^{-33}$  \\
1400 MHz flux density, $S_{1400}$ (mJy)\tablenotemark{b}        & 1.1  \\
2000 MHz flux density, $S_{2000}$ (mJy)\tablenotemark{b}        & 0.2  \\
$W_{50}$ at 1520 MHz (ms)\tablenotemark{c}                      & 0.20 \\
$W_{50}$ at 2000 MHz (ms)\tablenotemark{c}                      & 0.27 \\
$W_{50}$ at 3100 MHz (ms)\tablenotemark{c}                      & 0.21 \\
                                              &                                \\ 
\hline
Derived Quantities                            &                       \\
\hline \\
Spin period, $P$ (ms)                              & 1.855732795728(8)    \\  
Spin period derivative, $\dot{P}$                  & $7.5(4) \times 10^{-21}$ \\ 
Orbital period, $P_{b}$ (d)                   & 0.615436473(8)       \\	
Orbital period derivative, $\dot{P_{b}}$      & $-3.50(12) \times 10^{-9}$ \\
Companion mass range ($M_{\odot}$)\tablenotemark{d}          & 0.4$-$0.7             \\ 
Orbital inclination angle range (degrees)\tablenotemark{d}   & 30$-$41               \\ 
Galactic longitude, $l$ (deg)                 & 357.6  \\
Galactic latitude, $b$ (deg)                  & +4.3   \\ 
Surface magnetic field, $B$ (G)\tablenotemark{e} & $1.2 \times 10^{8}$ \\
Spin-down luminosity, $\dot{E}$ (erg s$^{-1}$)\tablenotemark{e} & $4.6 \times 10^{34}$   \\
Characteristic age, $\tau_{c}$  (Gyr)\tablenotemark{e} &  3.9 \\
Distance, $d$ (kpc)\tablenotemark{f}          & 0.75(10) \\
Distance from Galactic plane, $|z| = d \sin |b|$, (kpc) &  0.06 \\
1400 MHz radio luminosity, $L_{1400} = S_{1400} d^{2}$ (mJy kpc$^{2}$)  & $0.6$  \\
                                              &                              \\
\hline 
Observational Parameters                      &                       \\
\hline \\
Discovery observation MJD                     &  51973           \\ 
Phase-connected TOA range (MJD)               &  55102$-$55623   \\ 
Timing span (d)                               &  522             \\
Weighted rms post-fit residual  ($\mu$s)      & 20.9  \\ 
\enddata

\tablecomments{Figures in parentheses are uncertainties in the last
digit quoted and are twice the formal errors from the TEMPO timing
solution.}

\tablenotetext{a}{$x = a \sin i / c $ where $a$ is the semi-major axis
and $i$ is the orbital inclination angle.}

\tablenotetext{b}{Flux density measured away from eclipse. The value
was estimated using the radiometer equation applied to the original
survey discovery observation (for $S_{1400}$) and the GBT observation
shown in Figure \ref{fig-8} (for $S_{2000}$).}

\tablenotetext{c}{Width at which the measured pulse reaches 50\% of
its peak value. See Figure \ref{fig-8}.}

\tablenotetext{d}{Assumes a pulsar mass between 1.4 and 2.0 $M_{\odot}$.}

\tablenotetext{e}{$B = 3.2 \times 10^{19} (P \dot{P})^{1/2}$; $\dot{E}
= 4 \pi^{2} I \dot{P} / P^{3}$, with an assumed moment of inertia $I =
10^{45}$ g cm$^{2}$; $\tau_{c} = P / 2 \dot{P}$. These
parameters depend on $\dot{P}$ which has not been corrected for 
the Shklovskii effect.}

\tablenotetext{f}{Estimated from the NE2001 DM-distance model of
\citet{cl02}. The uncertainty in the distance was also 
obtained from this model.}

\end{deluxetable*}


The timing-derived period derivative is $\dot{P} = 7.5(4) \times
10^{-21}$. This not only includes Galactic acceleration and rotation
contributions, but also an undetermined contribution from the
Shklovskii effect \citep{s70}, which is caused by the pulsar's
transverse motion.  Following the analysis of \citet{nt95}, we find
that the acceleration and rotation contributions to $\dot{P}$ are
negligible ($< 2$\% combined). The Shklovskii effect, however, may be
significant given the small distance to the pulsar. The measured
$\dot{P}$ is highly sensitive to the transverse speed: a speed of 170
km s$^{-1}$ at the pulsar's estimated distance of 0.75 kpc would
produce a Shklovskii term that would entirely account for the measured
$\dot{P}$. This provides an upper limit of $\sim 170$ km s$^{-1}$
beyond which the Shklovskii $\dot{P}$ would exceed the measured
$\dot{P}$. Including a fit for proper motion in the timing solution is
unable to further constrain this upper limit owing to the relatively
large residuals in the solution and the limited timing span ($< 2$
years; see Table \ref{tbl-2}).

The spin-down power of the pulsar determined using the measured
(uncorrected) $\dot{P}$ is $\dot{E} = 4.6 \times 10^{34}$ erg s$^{-1}$
($\dot{E} \equiv 4 \pi^{2} I \dot{P} P^{-3}$, where $I$ is an assumed
neutron star moment of inertia of $10^{45}$ g cm$^{2}$).  The
$\dot{E}$ strongly affects the pulsar's interaction with its companion
and can produce ablation of the companion which obscures the pulsar's
signal. The measured $\dot{E}$ is typical of those measured for MSPs
listed in the ATNF Pulsar Catalogue, which range from $\sim 10^{32}$
to $\sim 10^{36}$ erg s$^{-1}$, and it is close to the values of
$\dot{E}$ measured for known Galactic redbacks, all of which have
$\dot{E} \sim 10^{34}$ erg s$^{-1}$ (see Deller et
al. 2012\nocite{dab+12} and Table 1 of Roberts
2013\nocite{r13}). However, none of these $\dot{E}$ measurements for
the redbacks (with the exception of PSR J1023+0038) have been
corrected for the Shklovskii effect, so the intrinsic $\dot{E}$ for
these pulsars could be much smaller.  The pulsar's characteristic age
is $\tau_{c} \equiv P/2\dot{P} > 3.9$ Gyr, and the estimated surface
magnetic field is $B \equiv (3.2 \times 10^{19}) (P \dot{P})^{1/2}
{\rm G} < 1.2 \times 10^{8}$ G. Again, these have not been corrected
for the Shklovskii effect, so lower and upper limits are quoted
here. These values are consistent with the values expected for a
recycled MSP.

\thepulsar\ follows an almost circular orbit, and the orbital
eccentricity is too small to be measured in our fit when the orbital
period derivatives are included in the timing solution.  The large
orbital period derivative (see Table \ref{tbl-2}) is a signature of a
strong tidal effect between the neutron star and an extended,
mass-losing companion \citep{lvt+11}. In addition, the pulsar is
eclipsed for a significant portion of its orbit ($\sim 15$\%; see
Figure \ref{fig-1}). The eclipsing occurs at orbital phases near 0.25,
when the pulsar is behind its companion. These features suggest that
the companion is a non-degenerate, extended star, which we confirm
below.

\begin{figure*}
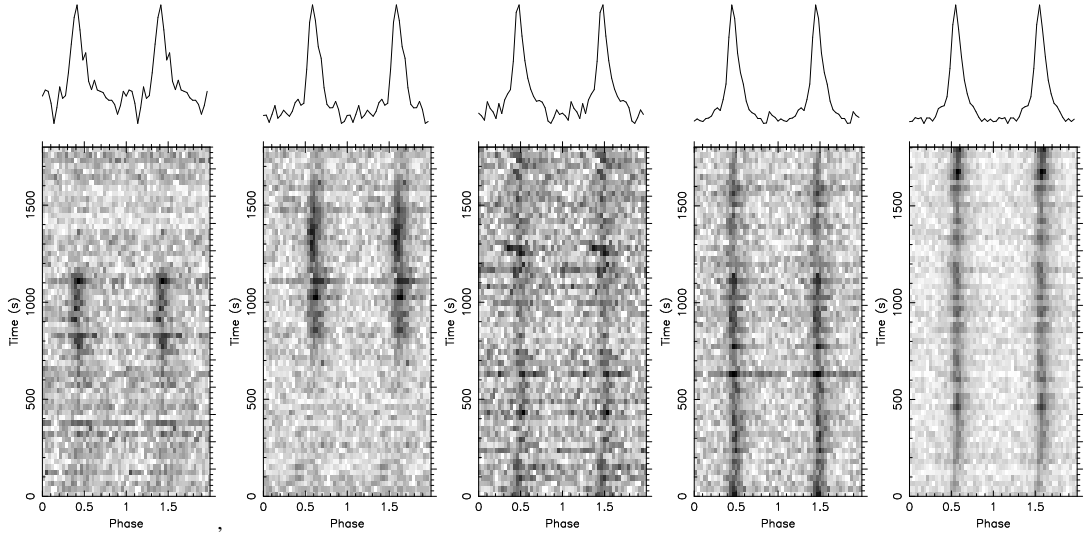

\centerline{\psfig{figure=fig2a.ps,width=1.1in,angle=270},
\psfig{figure=fig2b.ps,width=1.1in,angle=270}
\psfig{figure=fig2c.ps,width=1.1in,angle=270}
\psfig{figure=fig2d.ps,width=1.1in,angle=270}
\psfig{figure=fig2e.ps,width=1.1in,angle=270}}
\caption{Time-resolved folded pulse profiles of \thepulsar\ taken at
2000 MHz with the GBT. The time-integrated profile is shown above each
time-resolved profile plot. Each observation had an integration time
of 30 minutes. The flux variability is evident in these profiles and
is likely dominated by obscuration by material within the system and
not diffractive interstellar scintillation.\label{fig-2}}
\end{figure*}

\thepulsar\ exhibits significant flux variability on the time scale of
minutes.  Figure \ref{fig-2} shows detections at 2000 MHz with the GBT
which illustrate this variability.  One possible contribution to this
is diffractive scintillation, which may be significant given the
pulsar's small dispersion measure.  Dynamic spectra constructed using
both the Parkes 1.4 GHz discovery observation and a bright 2 GHz GBT
timing observation have decorrelation bandwidths of 12 MHz and 100
MHz, respectively, and decorrelation times of 18 min and 30 min. These
are all within a factor of two of the values predicted by the NE2001
model \citep{cl02}. The decorrelation times are similar to the
observed variability time-scales, but the decorrelation bandwidths are
much smaller in each case than the observing bandwidths used. This
suggests that eclipsing or obscuration by material from the companion
(and not scintillation) is the dominant factor in the variability.

\section{Photometric Identification and Spectroscopic Study of the Companion}

\subsection{Infrared, Optical, and Ultraviolet Photometry}

We identified the companion to \thepulsar\ in the infrared, optical,
and ultraviolet bands using photometry using both archival data and
new observations. This is outlined below and summarized in Table
\ref{tbl-3}.

\begin{deluxetable*}{l c c c c c}
\tablewidth{0pt}
\tablecaption{Photometry of the Optical Companion of \thepulsar\label{tbl-3}}
\tablehead{
\colhead{Band} & \colhead{Central} & \colhead{Measured} &
\colhead{Survey/Telescope\tablenotemark{a}} & \colhead{Observation} & \colhead{Magnitude\tablenotemark{b}}\\
 & \colhead{Wavelength} & \colhead{Magnitude} & \colhead{Source} & \colhead{Date} & \colhead{System} \\
 & \colhead{(\AA)}  
}
\startdata
UVW2  & \phn 1928 & $22.63 \pm 0.23$ & {\it Swift} & 2010 March 05     & AB   \\
$u$   & \phn 3450 & $18.53 \pm 0.05$ & {\it Swift} & 2010 March 05     & AB   \\
$B$   & \phn 4380 & $16.94 \pm 0.02$ & WIYN        & 2012 May 31       & Vega \\
$V$   & \phn 5450 & $15.78 \pm 0.04$ & WIYN        & 2012 May 31       & Vega \\
$R$   & \phn 6410 & $15.15 \pm 0.02$ & WIYN        & 2012 May 31       & Vega \\
$I$   & \phn 7980 & $14.42 \pm 0.07$ & WIYN        & 2012 May 31       & Vega \\
$Z$   & \phn 8780 & $14.25 \pm 0.01$ & VVV         & 2012 May 31       & Vega \\
$Y$   & 10210     & $14.07 \pm 0.01$ & VVV         & 2011 August 05    & Vega \\
$J$   & 12540     & $13.71 \pm 0.01$ & VVV         & 2011 August 05    & Vega \\
$J$   & 12350     & $13.65 \pm 0.05$ & 2MASS       & 1998 July 12      & Vega \\
$H$   & 16460     & $13.27 \pm 0.01$ & VVV         & 2010 August 03    & Vega \\
$H$   & 16620     & $13.21 \pm 0.02$ & 2MASS       & 1998 July 12      & Vega \\
$K_s$ & 21490     & $13.07 \pm 0.01$ & VVV         & 2011 September 22 & Vega \\
$K_s$ & 21590     & $13.07 \pm 0.02$ & 2MASS       & 1998 July 12      & ~~~Vega 
\enddata

\tablenotetext{a}{Survey or telescope origin of the measurement.
2MASS is the Two Micron All-Sky Survey \citep{2mass}.  VVV is the
VISTA Variables in the Via Lactea Survey \citep{mle+10}.}

\tablenotetext{b}{Measurements were in either the AB or Vega magnitude
system.}

\end{deluxetable*}


\subsubsection{Archival Optical and Near-Infrared Data} 

The timing position of \thepulsar\ places it within $0.06\arcsec$ of
the moderately bright Two Micron All-Sky Survey (2MASS;
\citealt{2mass}) star J17232318$-$2837571.  The number density of
stars brighter than this is 0.004 arcsec$^{-2}$, so there is a low
probability of coincidence and the association is secure.  The 2MASS
photometry is presented in Table \ref{tbl-3}.  Apart from 2MASS, the
source was clearly visible in the Digitized Sky Survey (DSS).
However, because of the poor quality of the DSS photometry we rely on
our new optical measurements, which are discussed below.  The star was
also detected in the Deep Near Infrared Survey of the Southern Sky
(DENIS; \citealt{edd+99}) with magnitudes Gunn-$i$ = $14.70 \pm 0.05$,
$J = 13.97 \pm 0.11$, and $K_s = 13.44 \pm 0.20$. The $J$ and $K_s$
magnitudes are about 0.3--0.4\,mag fainter than those from 2MASS.
While this might indicate variability, the crowded fields might also
have affected the automated photometry of one or both surveys.  Since
the 2MASS photometry is very close to other near-IR photometry (see
below and Table \ref{tbl-3}) we use those values, but urge caution in
making detailed inferences based on the precise numbers chosen.

\subsubsection{WIYN Optical Observations}

\begin{figure}
\centerline{\psfig{figure=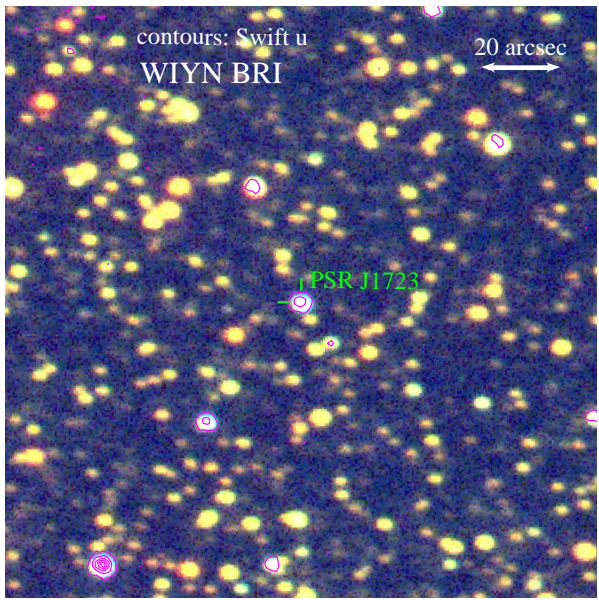,width=3.10in}}
\caption{Three-color composite image of the field of \thepulsar\
created from the WIYN 0.9-m data in the $B$, $R$, and $I$ filters.
Contours are from {\it Swift} $u$-band data.  The image is
$150\arcsec$ across, and North is up with East to the left.  The
position of \thepulsar\ is indicated by the tick marks.  The bar in
the upper right of the image is $20\arcsec$ in length.\label{fig-3}}
\end{figure}

We observed the companion of \thepulsar\ with the S2KB camera on the
Wisconsin Indiana Yale NOAO (WIYN) 0.9-m telescope on the night of
2012 May 31 under photometric conditions. We took two 300 s exposures
of the field in the $BVRI$ filters, along with two 180 s exposures of
the \citet{s00} field L111.  Reduction followed standard procedures in
IRAF.  We determined photometric zero-points from 20--150 stars in the
L111 field, depending on the band, with typical zero-point
uncertainties of 0.02 mag.  Our final photometry is given in
Table~\ref{tbl-3}.  We also show a three-color ($BRI$) composite image
of the field in Figure \ref{fig-3}, with contours from the {\it Swift}
$u$-band observation overlaid (see {\it Swift} observations below). It
is apparent that the optical companion to \thepulsar\ is brighter and
bluer than most of the field stars. We quantify this in the spectral
energy distribution in \S~\ref{sec:sed}.

\subsubsection{{\it Swift} Ultraviolet Observations}

An observation of the field with the {\it Swift} satellite from 2010
March 05 shows a bright source at the radio position in the data from
the Ultraviolet and Optical Telescope (UVOT; \citealt{rkm+05}).  In
the $u$ filter there was a single 765 s exposure, while in the UVW2
filter there were 13 exposures totaling 5577\,s; both bands used
$2\times 2$\,pixel binning.  We determined summed photometry from
these data using {\it Swift} data-reduction tools.  We summed the UVW2
integrations and measured the summed magnitude with a source radius of
$5\arcsec$ centered on the radio position and a background region
$25\arcsec$ in radius centered near the pulsar, but not including any
visible sources.  The final detection significances in the summed
images were 23.1\,$\sigma$ ($u$ filter) and 4.8\,$\sigma$ (UVW2
filter), but our photometry includes the suggested systematic
uncertainties (0.02 mag and 0.03 mag, respectively) in addition to the
statistical uncertainty.  The details of the photometry are presented
in Table \ref{tbl-3}.

\subsubsection{VVV Infrared Data}

Finally, we assembled photometry from the VISTA Variables in the Via
Lactea Survey (VVV; \citealt{mle+10}) Data Release 1.  Although this
is a time-domain survey, we only had access to a single epoch of data
for each filter ($Z Y J H K_{s}$).  We followed the recommended
procedure to extract calibrated photometry from the data\footnote{See
\url{http://www.eso.org/sci/observing/phase3/data\_releases/vvv\_dr1.html}.},
using the recommended photometric zero-point uncertainty of 0.01\,mag.
We note that while the VVV photometry has been calibrated with respect
to 2MASS, \citet{shm+12} gives slightly different effective
wavelengths than \citet{cwm03}.  Our final photometry is given in
Table~\ref{tbl-3}.

\subsection{Optical Spectroscopy with Palomar and WIYN}

After the photometric identification of the coincident star, we
obtained two epochs of low-resolution spectroscopy with two
telescopes. We observed the companion of \thepulsar\ with the Double
Spectrograph (DBSP) on the 5-m Palomar (Hale) telescope. We obtained
two 300 s exposures on 2012 March 20 and 21.  The blue side used the
$600\,{\rm line\,mm}^{-1}$ grating (1.07 \AA\ pixel$^{-1}$, covering
3700-5500\,\AA) while the red side used the $316\,{\rm line\,mm}^{-1}$
grating (1.53 \AA\ pixel$^{-1}$, covering 5500-10000\,\AA, although
beyond 7000\,\AA\ fringing and uncertain telluric corrections make the
data not very useful), with the 5500\,\AA\ dichroic separating the
sides and a $1\arcsec$ slit.  Reduction followed standard procedures
in IRAF.  We determined the wavelength solution using Fe/Ar arc lines
for the blue side (61 lines with an rms residual of $0.22\,$\AA) and
He/Ne/Ar lines for the red side (27 lines with an rms residual of
$0.07\,$\AA).  Rough flux calibration was relative to the
spectrophotometric standard star BD+28\,4211.

We also observed the companion of \thepulsar\ for two 300 s exposures
on 2012 March 21 and 22 using the Bench Spectrograph with the
Sparsepak fiber bundle \citep{bah+04} on the 3.5-m WIYN telescope.
These observations used the Red camera and the $600\,{\rm
line\,mm}^{-1}$ grating, centered at a wavelength of 5200\,\AA, with
the CCD binned substantially (4 pixels in the spatial dimension and 3
in the spectral dimension).  Reduction followed standard
procedures\footnote{See
\url{http://www.astro.wisc.edu/$\sim$cigan/reducing/reducing.html}.}
in IRAF.  We determined a wavelength solution using 49 lines from the
Cu/Ar lamp, with an rms residual of $0.7\,$\AA.  The background was
subtracted using a nearby fiber that was determined not to cover a
star (as identified from the images discussed above).  The wavelength
coverage was 3800-6630 \AA\ with a scale of $2.1$ \AA\ pixel$^{-1}$.
Rough flux calibration was done with an observation of the
spectrophotometric standard HZ44.

Table \ref{tbl-4} summarizes these two sets of spectral observations.
For both sets the pulsar was at an airmass of between 2.0 and 2.2, so
the flux calibration and telluric line removal in the red may not be
very reliable.

\begin{deluxetable}{lcccc}
\tablecaption{Doppler Spectroscopy of the Optical Companion of \thepulsar\label{tbl-4}}
\tablewidth{0pt}
\tablehead{
\colhead{Telescope} &
\colhead{Date (UT Time)} &
\colhead{MJD} &
\colhead{Velocity} &
\colhead{Orbital} \\
\colhead{} &
\colhead{} &
\colhead{} &
\colhead{(km s$^{-1}$)} &
\colhead{Phase} 
}
\startdata
Palomar & 2012 Mar 20 (12:41) & 56006.52852 & $+145 \pm 5$   & 0.39 \\
WIYN    & 2012 Mar 21 (12:28) & 56007.51978 & $-52  \pm 25$  & 1.00 \\
Palomar & 2012 Mar 21 (12:59) & 56007.54095 & $-116 \pm 10$  & 0.03 \\
WIYN    & 2012 Mar 22 (11:57) & 56008.49839 & $+127 \pm 15$  & ~~0.59 
\enddata

\tablecomments{Times and velocities have been converted to the Solar System
barycenter. Times correspond to the middle of the 300 s exposure in each
case.}

\end{deluxetable}
				   	     

\subsection{Spectral Type and Radial Velocities}

\label{sec:rv}
\begin{figure}
\centerline{\psfig{figure=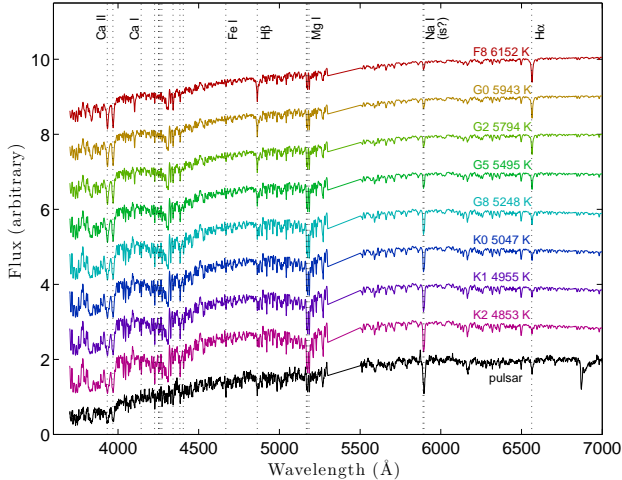,width=3.25in}}
\caption{Spectrum of the optical companion to \thepulsar\ from the
2012 March 20 Palomar observation (bottom), along with main-sequence
model spectra from \citet{mscz05}.  The models go (top to bottom) from
F8 (effective temperature 6150 K) to K2 (4850 K), as labeled. The
model spectra have been convolved to match the resolution of the data
and have been multiplied by a linear function to match the continuum
slope of the data.  Selected absorption lines are labeled, with the
possible interstellar contribution to the \ion{Na}{1} line
indicated.\label{fig-4}}
\end{figure}

\begin{figure}
\centerline{\psfig{figure=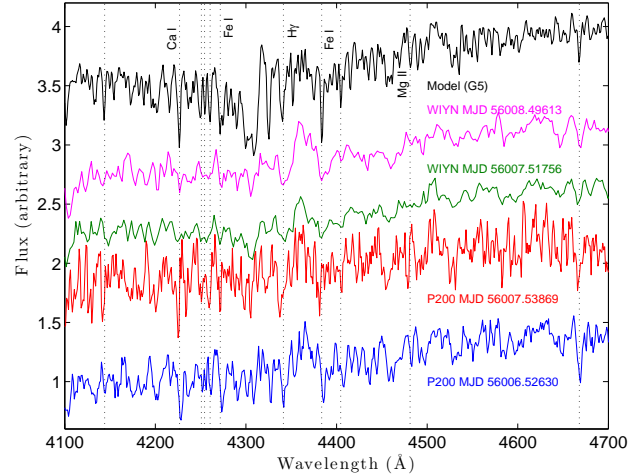,width=3.25in}}
\caption{A G5 model spectrum is shown (at the top) along with the
measured spectra of the optical companion to \thepulsar\ from the
Palomar and WIYN observations (labeled).  Selected absorption lines
are also labeled.  The velocity shift between the two Palomar spectra
is readily visible (e.g., in the \ion{Ca}{1} $\lambda 4226$ line),
although similar shifts are also seen in the WIYN spectra. The MJD
values shown have not been corrected to the Solar System barycenter.
The measured Doppler velocities from these spectra are presented in
Table \ref{tbl-4}.\label{fig-5}}
\end{figure}

We used our spectra to try to establish a spectral type for the
optical companion.  We show one of the Palomar spectra in
Figure~\ref{fig-4}, along with model spectra of types F8-K2 (see also
Figure \ref{fig-5}). A comparison of our spectra to these model
spectra as well as to the Gray spectral atlas\footnote{See
\url{http://ned.ipac.caltech.edu/level5/Gray/frames.html}.} indicates
that the spectral type of the companion seems to be somewhere around
FGK, although the precise type is hard to pin down.  The \ion{Na}{1}
$\lambda 5891$ line is strong, as in a late G or early K star, but
some of this could be from interstellar absorption; a full sampling of
radial velocities should be able to disentangle this.  This appears to
agree with the modest depth of the \ion{Ca}{2} H and K lines.
\ion{Cr}{1} $\lambda$5254 is about equal in strength to \ion{Fe}{1}
$\lambda $4250 and $\lambda 4260$, which would place the companion
somewhere near K0.  \ion{Fe}{1} $\lambda$4325 is roughly equal in
depth to H$\gamma$, which would put it somewhere between G5 and K0.
However, the molecular G-band is rather weak, more similar to an
earlier G star.  Other lines like \ion{Fe}{1} $\lambda 4668$ also
appear stronger in the observed spectrum than in all of the models
(other \ion{Fe}{1} lines are similarly strong).  The \ion{Mg}{1} lines
near 5180\,\AA\ appear modest in strength, but the MgH complex that
overlaps with those lines in later-type stars is not apparent.
Overall, our spectral data indicate that the optical companion to
\thepulsar\ has an effective temperature of 5000-6000 K and a surface
gravity that is reasonably consistent with that of a main-sequence
star. Beyond that we cannot make firm conclusions from these spectra.
It could be that the metallicity of the companion is significantly
different from the templates or that some other effect is causing
variations in individual lines (e.g., rotation or an anomalously low
surface gravity). The pulsar's wind also has a significant heating
effect on the companion's atmosphere, which may modify the apparent
spectral type and would contribute to its uncertainty.

\begin{figure}
\centerline{\psfig{figure=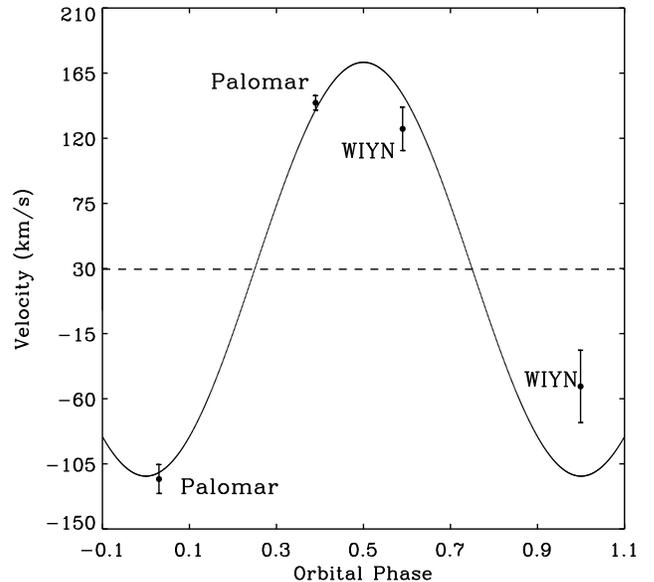,width=3.5in}}
\caption{Measured Doppler velocities from the Palomar and WIYN spectra
(see Table \ref{tbl-4}). The velocities have been corrected to the
Solar System barycenter. Overlaid is the best-fit sinusoid with a
fixed phase, period, and offset (taken from the barycenter
correction), but with an unknown amplitude.  The best fit velocity
amplitude is 143 km s$^{-1}$ with an estimated uncertainty of $\pm 20$
km s$^{-1}$. For an assumed pulsar mass range of 1.4$-$2.0
$M_{\odot}$, the corresponding companion mass range is 0.4$-$0.7
$M_{\odot}$, and the orbital inclination angle is between 30 and
$41^{\circ}$.\label{fig-6}}
\end{figure}

We then determined the radial velocities of the individual
observations.  We did this by fitting each spectrum to a shifted and
convolved template of a type G5 star.  For the Palomar data, we
verified that the sky lines at 4358\,\AA, 5460\,\AA, and 5577\,\AA\
were consistent between the two observations.  For the WIYN data, the
4358\,\AA\ sky line was used to establish the velocity reference, and
we used the data between 3880\,\AA\ and 5000\,\AA\ (where we had good
sensitivity in all observations).  We determined velocities for both
Palomar observations and for both WIYN observations and corrected the
observed velocities to the Solar System barycenter.  These
measurements are presented in Table \ref{tbl-4} and are shown in
Figure \ref{fig-6}. A region of each spectrum is shown in Figure
\ref{fig-5}, where the shift between the Palomar observations is
readily apparent.  The lower resolution and signal-to-noise ratio of
the WIYN observations make the shift harder to discern, but it is also
present.

The Palomar and WIYN observations from 2012 March 21 were taken 30 min
apart from each other and are at almost the same orbital phase.  While
the velocities are not formally consistent with each other within the
uncertainties (see Table \ref{tbl-4}), the uncertainties of the WIYN
observations may be somewhat underestimated, and the disagreement is
considerably less than the change between the two Palomar
observations.  Moreover, the measured velocities are consistent with
those expected at the various orbital phases.  This is additional
evidence supporting the association between the star and \thepulsar.
Figure \ref{fig-6} shows a sinusoidal fit (with fixed phase, period,
and offset) to the four measured velocities. The best-fit velocity
amplitude is 143 km s$^{-1}$, with an uncertainty estimated to be $\pm
20$ km s$^{-1}$ owing to the meager sampling and inhomogeneous data.
Using this measured velocity amplitude and the projected semi-major
axis for the pulsar (Table \ref{tbl-2}), we derive a mass ratio of
$3.3 \pm 0.5$, giving a companion mass range of 0.4 to 0.7 $M_{\odot}$
and an orbital inclination angle range of 30 to $41^{\circ}$ for an
assumed pulsar mass of between 1.4 and 2.0 $M_{\odot}$. This pulsar
mass range reflects the fact that some MSPs have measured masses above
the canonical value of 1.4 $M_{\odot}$ (e.g., Demorest et
al. 2010\nocite{dpr+10}).  The inhomogeneity of the data and the small
number of observations preclude a more precise radial-velocity
measurement, but further observations will be able to easily determine
this number to high precision and may allow for an independent
estimate of the pulsar mass if the inclination angle can be
independently constrained via light curve modeling (e.g., van Kerkwijk
et al. 2011\nocite{vbk11}; Romani et al. 2012\nocite{rfs+12}).

\subsection{Spectral Energy Distribution}

\label{sec:sed}
\begin{figure}
\centerline{\psfig{figure=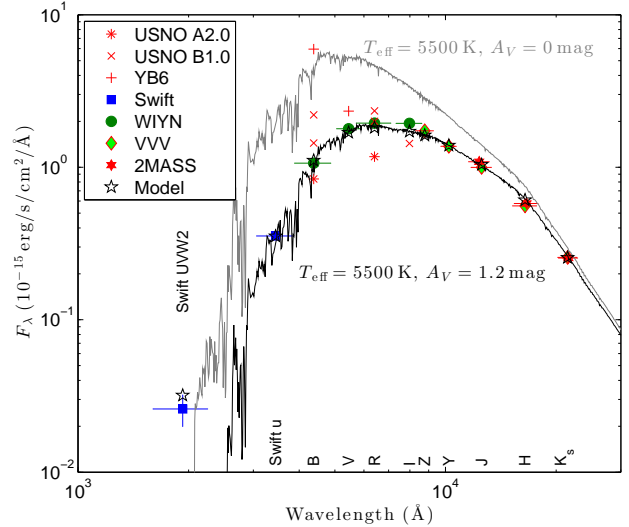,width=3.25in}}
\caption{Spectral energy distribution of the optical companion to
\thepulsar.  The data from {\it Swift} (squares), WIYN (circles), VVV
(diamonds), and 2MASS (stars) surveys are plotted along with the
best-fit spectral model from \citet{kurucz93}.  The black trace is a
5500 K (roughly G5) model including an extinction of $A_V = 1.2$ mag,
while the gray trace is without extinction; the black stars represent
the black trace model convolved with the filter response curves.  We
also plot data from the USNO-A2.0, USNO-B1.0, and YB6 catalogs
\citep{monet+98,mlc+03} based on the Naval Observatory Merged
Astrometric Dataset (NOMAD).\label{fig-7}}
\end{figure}

With the range of spectral types established in \S~\ref{sec:rv} and
Figure \ref{fig-4}, we can see whether the magnitudes from photometry
are consistent with the spectral values, and we can use these
magnitudes to determine the normalization and extinction.  Using the
best photometry available (the {\it Swift} $u$-band, WIYN, and VVV
data), we fit reddened models from \citet[][since the models from
\citealt{mscz05} did not extend into the near-infrared]{kurucz93}.  We
obtained reasonable fits for models consistent with our spectral
inferences: effective temperatures between 4800 K and 6000 K all fit,
with the extinction varying between $A_V = 0.2$ mag to 1.7 mag over
that range.  We did have to assume a modest (5\%) systematic
uncertainty, but given the variations in the assumed filter
transmission curves and the lack of color terms in our photometric
calibration, this was reasonable.  The best fit was for an effective
temperature of 5500 K (spectral type G5) and an extinction of $A_V =
1.2$ mag, as shown in Figure~\ref{fig-7}, although the exact fit
varies a little if we change the systematic uncertainties.  This
temperature is somewhat higher than the temperature expected from the
heating of the companion from the pulsar wind if 100\% of the measured
$\dot{P}$ were intrinsic to the pulsar, if the wind were isotropic,
and if the companion had a 15\% absorption efficiency factor (see,
e.g., Breton et al. 2013\nocite{bvr+13}). The expected temperature
range from this would be 3100-3600 K for the inclination angle range
30 to $41^{\circ}$.  While we did not use the 2MASS or {\it Swift}
UVW2 data for the fitting, our model also fits these data well (the
2MASS data are largely consistent with the VVV data).  We also see
that the scanned-plate data from the USNO-A2.0, USNO-B1.0, and YB6
catalogs \citep{monet+98,mlc+03} are plausibly matched by the data.
There is some scatter, and this may arise from difficulties in
converting the USNO photometry to fluxes, or it could indicate
variability in the photometry across the orbit, as in the case of PSR
J1023+0038 \citep{wat+09}.  Based on the observed $K_s$-band flux, we
estimate a normalization of $1.2\,R_{\odot}\,{\rm kpc}^{-1}$.

\section{Discussion}

\begin{figure}
\centerline{\psfig{figure=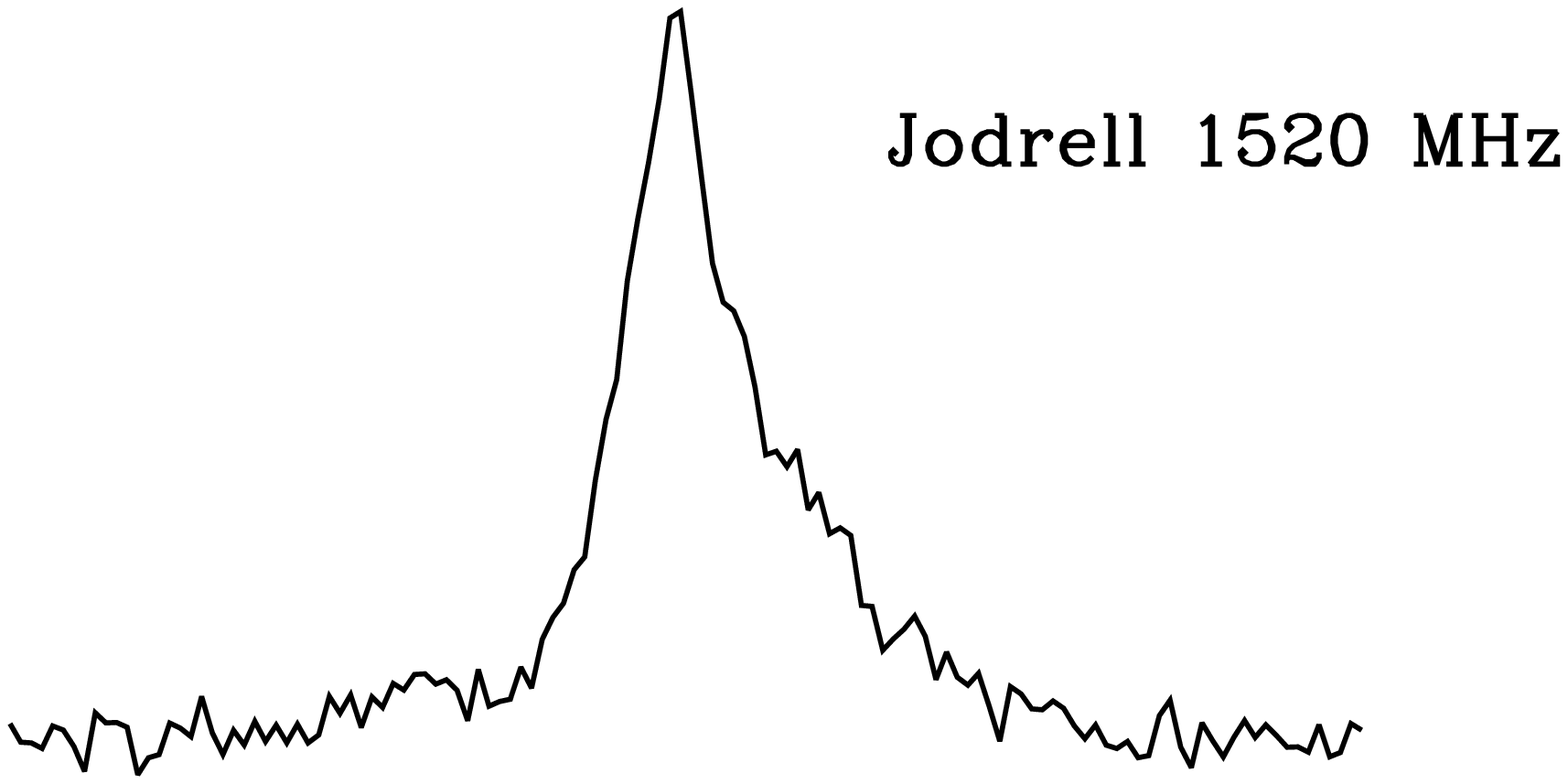,width=2in}}
\centerline{\psfig{figure=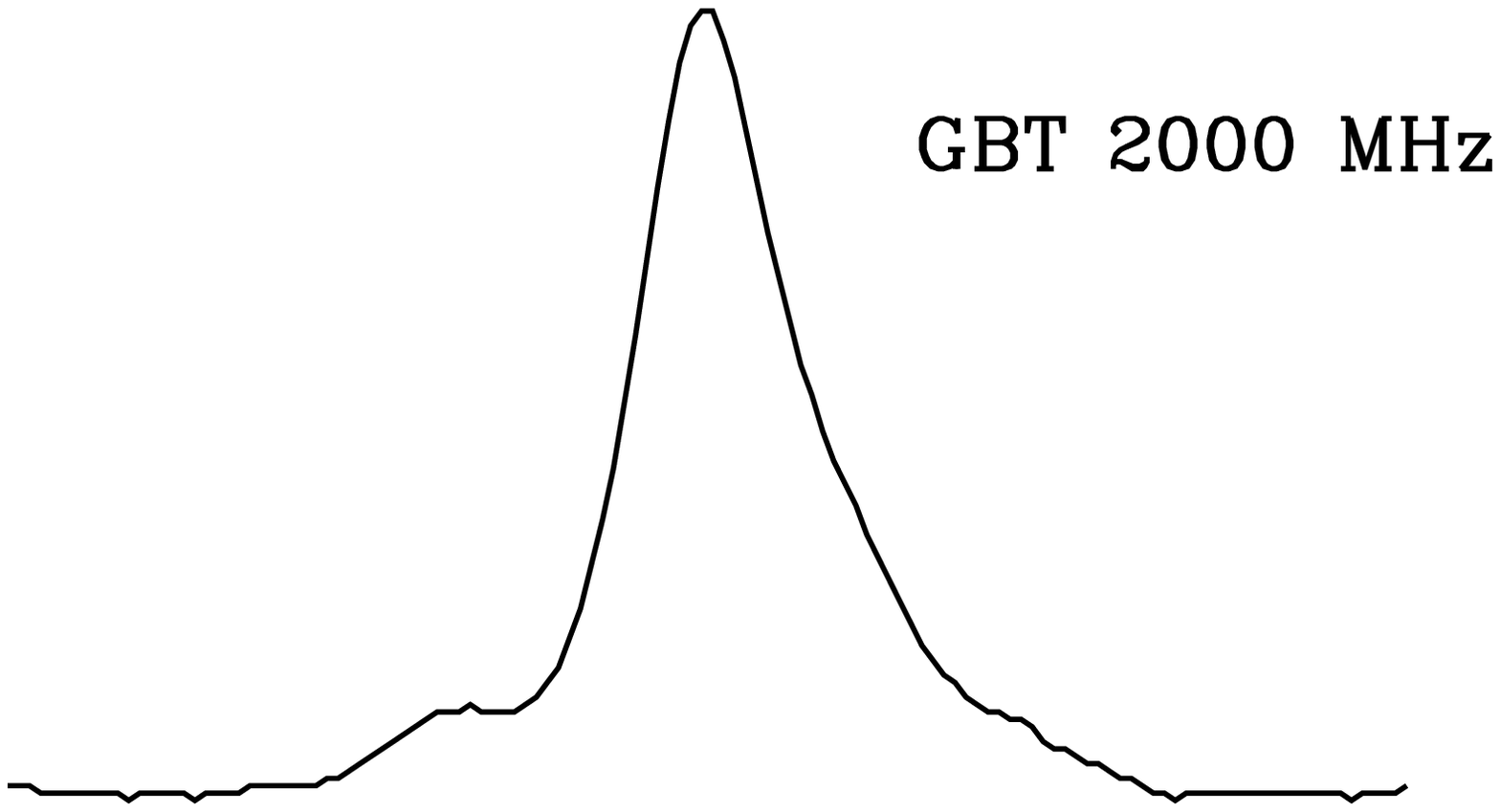,width=2in}}
\centerline{\psfig{figure=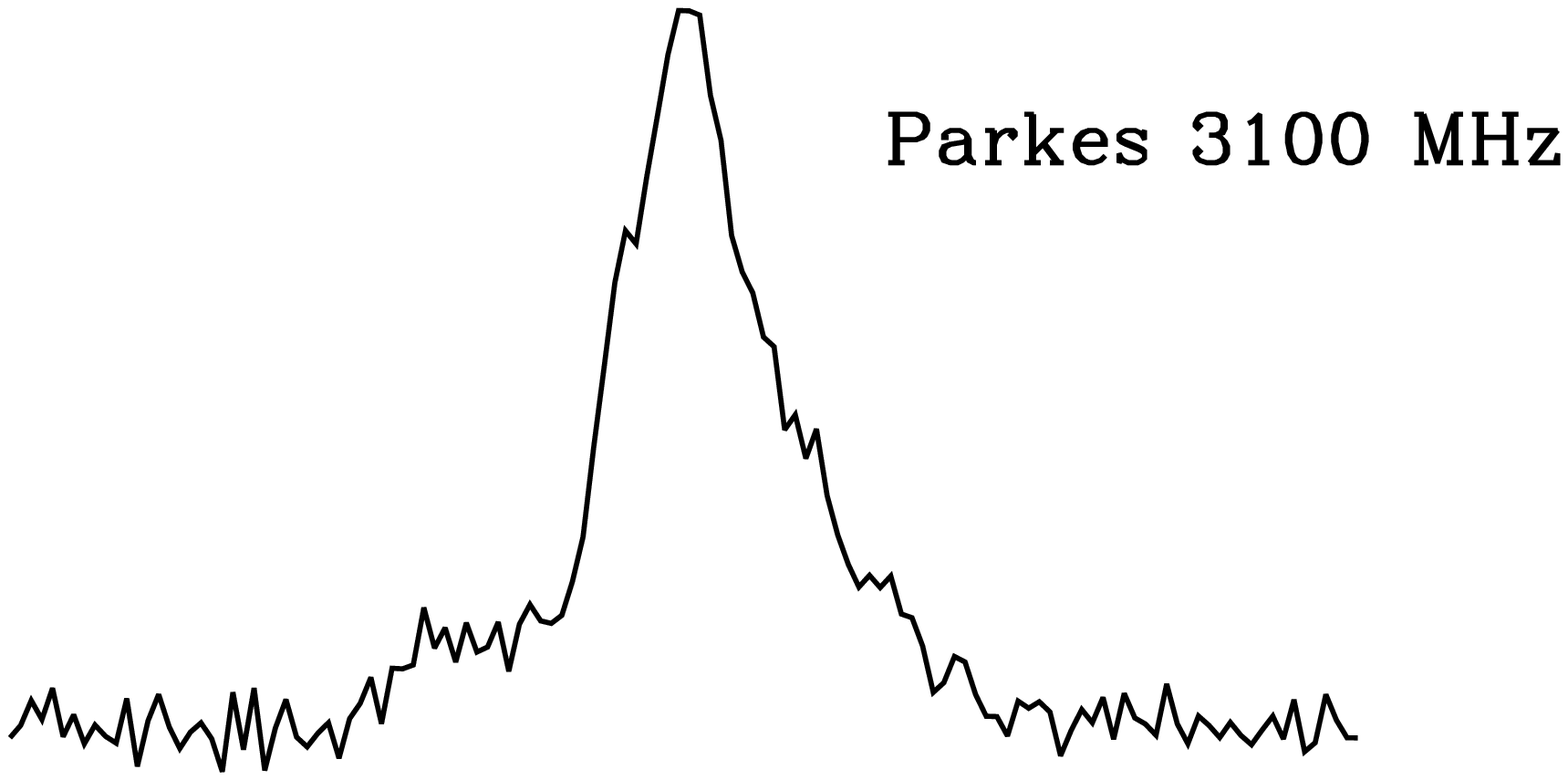,width=2in}}
\caption{Pulse profiles for \thepulsar\ at 1520 MHz from Jodrell Bank
(top), 2000 MHz from the GBT (middle), and 3100 MHz from Parkes
(bottom).  Each profile has 128 bins, and the full profile is shown in
each case.  The profiles were constructed from observations of 40, 62,
and 60 minutes, respectively.  In all cases the observations were
taken at orbital phases far from the eclipsing region.  The similarity
between the profiles indicates that the trailing extension of the
profile is not due to interstellar scattering.\label{fig-8}}
\end{figure}

Figure \ref{fig-8} shows integrated pulse profiles for \thepulsar\ at
1520, 2000, and 3100 MHz that were taken at orbital phases far from
eclipse.  There is evidence of asymmetry in all of the profiles, which
could indicate interstellar scattering.  However, several pieces of
evidence argue against this.  First, the NE2001 model indicates that
the scattering time-scale from the interstellar medium in this
direction for a DM of 20 pc cm$^{-3}$ is negligible ($\la 10^{-7}$ s
at 1 GHz).  Second, the scatter-broadening time would be significantly
larger at 1520 MHz than at 3100 MHz (by a factor of $\sim 11$ for a
power-law frequency dependence of $\nu^{-3.4}$; L{\"o}hmer et
al. 2001\nocite{lkm+01}). As seen in Figure \ref{fig-8}, the profiles
at all frequencies appear to have a similar tail length, which would
not be expected from scattering. We conclude that the asymmetry in the
pulse profiles is intrinsic to the beam shape and that the profile
does not evolve significantly between these frequencies.

It is clear that eclipsing by the companion introduces a large flux
variability throughout the orbit at all observed frequencies, and this
is responsible for some of our non-detections at 1400 MHz with Parkes.
Figure \ref{fig-1} shows the timing residuals as a function of orbital
phase. There are no detections for $\sim 15$\% of the orbit, centered
near a phase of 0.25, when the pulsar is behind the companion (at
inferior conjunction).  This is partially due to the scheduling of
most of the observations to occur away from the eclipse phase range in
order to maximize the likelihood of detection.  However, some of the
2000 MHz GBT observations were taken at orbital phases between 0.5 and
0.6, well away from the eclipse phases near 0.25, and the pulsar was
not detected in these cases.  This suggests that eclipsing may not be
the only reason for the non-detections; for instance, the pulsar might
be shrouded at times by gas or an extended companion wind that is
being driven by the pulsar (see below).  \thepulsar\ is similar in
this sense to the GC pulsars PSR J0024$-$7204W
(47~Tuc~W) and PSR J1740$-$5340. PSR J0024$-$7204W has a low-mass
companion and exhibits eclipses lasting for about half of the orbital
period \citep{clf+00,egc+02}.  PSR J1740$-$5340 is an eclipsing pulsar
with a bright optical companion that fills its Roche lobe
\citep{dpm+01, fpd+01}. There are other examples of such pulsars in
GCs: PSR J1701$-$3006B in M62 \citep{pdm+03}, which has a companion
optically identified by \citet{cfp+08}, and PSRs J1824$-$2452H and
J1824$-$2452I in M28 \citep{pdf+10, pfb+13}.

The sparse orbital coverage of the available photometric data does not
allow us to constrain the degree of tidal distortion of the companion
star. However, the similarity of \thepulsar\ to both PSR J1023$+$0038
\citep{asr+09} and PSR J1740$-$5340 \citep{fpd+01} suggests that the
companion could be a star nearly filling its Roche lobe.  For an
orbital geometry using reference masses of 0.43 and $1.4 M_{\odot}$
for the companion and the pulsar, respectively, the photometric data
imply a distance of 0.77 kpc for the star. This nicely agrees with the
0.75 kpc distance estimated for the pulsar from the DM.

For a pulsar mass range 1.4$-$2.0 $M_{\odot}$, the corresponding
radius for a Roche-lobe filling companion star is $\sim 1$ $R_{\odot}$
(e.g., Breton et al. 2013\nocite{bvr+13}). As expected, this radius is
considerably larger than the radius of a star with the mass of the
companion, suggesting substantial bloating of the companion, as is
seen in the J1023$+$0038 and J1740$-$5340 systems. This radius would
give an eclipse fraction of $\sim 9$\% of the orbit if seen edge-on,
in contrast to the observed eclipse fraction of $\sim 15$\%, which
corresponds to a eclipsing radius of $\sim 1.7$ $R_{\odot}$.  This
eclipsing radius becomes even larger when the inclination angle is
taken into account.  This is an indication that the extended eclipse
range is likely caused by a plasma wind or a cloud of outflowing
matter released from the companion which overflows the Roche lobe and
thus must be continuously replenished.

The features of \thepulsar\ suggest that it is a member of the growing
class of redback systems.  Redbacks are binary millisecond radio
pulsars with short-period, circularized orbits and with extended
companions having masses of a few tenths of a solar mass (see Roberts
2013\nocite{r13}).  As mentioned above, the first two objects likely
belonging to this class were found about a decade ago in GCs.  The
first system in the Galaxy to be confirmed as a redback, PSR
J1023+0038 \citep{asr+09}, is believed to have turned on as a radio
pulsar after a recent phase in which there was an accretion disk which
is no longer present in the system \citep{wat+09}. This pulsar is
thought now to be in a bi-stable state, switching between radio pulsar
and LMXB phases according to the changing rate at which the companion
overflows its Roche lobe, as is observed for PSR~J1824$-$2452I
\citep{pfb+13}.  All of the currently known Galactic redbacks
(including \thepulsar) are nearby field pulsars ($d \la 3$ kpc).

Given that all Galactic redbacks discovered so far have been
successfully detected as $\gamma$-ray sources with {\it Fermi}, it is
likely that there are high-energy counterparts to \thepulsar. The
SIMBAD database\footnote{http://simbad.u-strasbg.fr/simbad/} indicates
that there is a cataloged {\it ROSAT} X-ray point source
(\rosatsource) located $13''$ from the position of \thepulsar. The
uncertainty in the X-ray position is given as $17''$, making its
position consistent with that of \thepulsar. The average sky density
of {\it ROSAT} sources in the 1RXS catalog \citep{vab+99} is low (less
than one cataloged source per square degree), making the likelihood of
a chance coincidence with the pulsar very small ($\sim 10^{-5}$).
A cataloged {\it INTEGRAL} hard X-ray source (\integralsource) is also
located only $\sim 1'$ from the pulsar \citep{ktr+10}. This is also a
likely association given the large uncertainty of the source position.
In contrast, the second Fermi LAT catalog (2FGL) \citep{2fgl} shows no
cataloged source within one degree of \thepulsar.

As discussed by \citet{r13}, there are five other redback systems with
characteristics like \thepulsar, all of which are Fermi-detected
sources. Table \ref{tbl-5} lists these five systems and \thepulsar\
ranked by $\dot{E} / d^{2}$, which is a measure of detectability at
high-energies. $\dot{E}$ and $d$ in this table were obtained from
Table 1 of \citet{r13}, but only PSR J1023+0038 has an $\dot{E}$ that
has been corrected for the Shklovskii effect and for Galactic
accelerations \citep{dab+12}. As seen in Table \ref{tbl-5},
\thepulsar\ ranks first among this group, indicating that high-energy
emission ought to be detectable.  However, a fold of {\it Fermi}
photons with the pulsar ephemeris using a variety of energy and angle
cuts shows no clear evidence of pulsed emission.  This may be due in
part to the large background in this part of the sky (near the
Galactic plane), or \thepulsar\ might be sub-luminous in $\gamma$-rays
owing to the orientation of the emission geometry, as described by
\citet{rkc+11}.  The proper motion might also be near the upper limit
of 170 km s$^{-1}$ obtained from the measured $\dot{P}$ and pulsar
distance. In this case, the intrinsic $\dot{P}$ and $\dot{E}$ would be
much smaller than the measured values. For a proper motion of $\sim
30$ km $s^{-1}$, which is comparable to the heliocentric radial speed
of the system, the motion would contribute only a few percent of the
measured $\dot{P}$ and would not significantly affect the measured
values.

We can compute an upper limit to the $\gamma$-ray efficiency for
\thepulsar\ by considering the minimum point-source fluxes for sources
in the Galactic plane in the 2FGL catalog \citep{2fgl}. For this we
assume an upper limit of $5 \times 10^{-12}$~erg~cm$^{-2}$~s$^{-1}$ on
the 100 MeV - 100 GeV flux. Assuming that the entire measured
$\dot{P}$ is intrinsic to \thepulsar, the $\gamma$-ray efficiency is
0.007, which is lower than the typical $\gamma$-ray efficiencies of
$\sim 0.1$ for MSPs. We therefore adopt this as an upper limit on the
$\gamma$-ray efficiency. This value is lower than the values measured
for the Galactic redbacks listed in Table \ref{tbl-5}, in some cases
by more than an order of magnitude.

\begin{deluxetable}{lcccc}
\tablecaption{Galactic Redbacks \label{tbl-5}}
\tablehead{
\colhead{Pulsar} &
\colhead{$\dot{E}_{34} / d_{kpc}^{2}$} &
\colhead{Reference} &
\colhead{{\it Fermi}} &
\colhead{$\gamma$-ray\tablenotemark{a}} \\
\colhead{} &
\colhead{} &
\colhead{} &
\colhead{Detected?} &
\colhead{Efficiency}
}
\startdata
J1723$-$2837 & 8.2                  & this paper     & no  & $< 0.007$ \\ 
J2129$-$0429 & 4.8                  & \citet{hrm+11} & yes & 0.02 \\ 
J1023$+$0038 & 2.3                  & \citet{thh+10} & yes & 0.03 \\
J1628$-$3205 & 1.3                  & \citet{rap+12} & yes & 0.11 \\
J1816$+$4510\tablenotemark{b} & 0.9 & \citet{ksr+12} & yes & 0.20 \\
J2215$+$5135 & 0.7                  & \citet{hrm+11} & yes & ~~~~0.17 
\enddata

\tablecomments{Entries are rank-ordered by $\dot{E_{34}} /
d_{kpc}^{2}$, an indicator of detectability at high
energies. $\dot{E_{34}} = \dot{E} / 10^{34}$ erg s$^{-1}$ and
$d_{kpc}$ is the distance in kpc. $\dot{E}$ values should be
considered upper limits in all cases (except for J1023$+$0038) since
the measured $\dot{P}$ has not been corrected for proper motion.
Values of $\dot{E}$ and $d$ for all pulsars
(except for \thepulsar\ and J1023$+$0038) were taken from Table 1 of
\citet{r13}.}

\tablenotetext{a}{Assumes that the measured $\dot{P}$ is entirely
intrinsic to the pulsar.}

\tablenotetext{b}{The companion to PSR J1816$+$4510 appears to be a
white-dwarf-like object with substantial metals \citep{kbv+13} and is
therefore unlike the other redbacks listed.}

\end{deluxetable}

\section{Conclusions}

We present a study of the binary radio MSP \thepulsar. We have
determined a phase-connected timing solution for the pulsar, which is
presented in Table \ref{tbl-2}. It is evident that the pulsar has a
low-mass, extended companion and is in a circularized, short-period
orbit.  We have identified the pulsar's companion using infrared,
optical, and ultraviolet photometry and spectroscopy. The mass ratio
of the system measured from Doppler variations in the companion star's
spectrum is $3.3 \pm 0.5$ (ratio of pulsar to companion mass). For an
assumed pulsar mass range of 1.4$-$2.0 $M_{\odot}$, the corresponding
companion mass range is 0.4 to 0.7 $M_{\odot}$ and the orbital
inclination angle range is 30 to $41^{\circ}$.  The best-fit spectrum
for the companion indicates that it has a spectral type of G
(effective temperature near 5500 K), which is consistent with the
measured spectral energy distribution.  The distance inferred from the
observations of the companion is also consistent with the estimate
obtained from the DM of the pulsar. However, the stellar radius of the
companion is larger than expected, indicating that the star is likely
close to filling its Roche lobe.  The character of the flux
variability of the pulsar, the difficulty in detecting it at low
frequencies, and the eclipse fraction of $\sim 15$\% (much larger than
the 9\% expected from the filled Roche lobe of the companion if it
were seen edge-on) also indicates that there is likely an extended
region of stellar plasma responsible for the eclipsing. Measurements
of the scintillation times and bandwidths indicate that diffractive
scintillation does not play a dominant role in the observed flux
variability.  The pulsar's large spin frequency indicates that it is
highly recycled (in the standard MSP production model) despite having
a companion that superficially resembles a main-sequence star.  These
features place it in the category of Galactic redbacks \citep{r13}.
In some respects this system is similar to PSR J1023+0038 and supports
the standard evolutionary picture of radio MSPs \citep{acr+82}.
Unlike the other five Galactic redback systems discovered to date,
\thepulsar\ is not detected as a {\it Fermi} source. There are several
possibilities for why it is not detected. A Shklovskii correction to
$\dot{E}$ from proper motion has not been accounted for, and the
pulsar is located near the Galactic plane, where the background is
brighter. The pulsar might also be sub-luminous in $\gamma$-rays owing
to its emission orientation.  \thepulsar\ is coincident with a {\it
ROSAT} source and a cataloged {\it INTEGRAL} hard X-ray source, both
of which are likely associated with the pulsar.  Future radio
polarimetry observations might be able to detect rotation measure
variations as a function of orbital phase, which could be useful for
probing the magnetized stellar wind from the companion star.

\acknowledgements 

We thank David Levitan, Marsha Wolf, and Eric Hooper for assistance
with the WIYN observations, and we thank Claire Gilpin and Deborah
Schmidt for assistance with radio timing observations with the GBT.
We also thank the anonymous referee for helpful suggestions which have
improved the manuscript.  The National Radio Astronomy Observatory is
a facility of the National Science Foundation operated under
cooperative agreement by Associated Universities, Inc.  The Parkes
radio telescope is part of the Australia Telescope, which is funded by
the Commonwealth of Australia for operation as a National Facility
managed by CSIRO. STScI is operated by the Association of Universities
for Research in Astronomy, Inc., under NASA contract NAS5-26555.  This
research was supported in part by NRAO Student Observing Support Award
GSSP09-0006. Pulsar research at UBC is supported by an NSERC Discovery
Grant.  P.F. gratefully acknowledges the financial support provided by
the European Research Council for the ERC Starting Grant BEACON under
contract no. 279702. Part of this work was based on data products from
observations made with ESO Telescopes at the La Silla or Paranal
Observatories under ESO programme ID 179.B-2002, and from the WIYN
Observatory, which is a joint facility of the University of
Wisconsin-Madison, Indiana University, Yale University, and the
National Optical Astronomy Observatories.  Some of the data presented
were obtained from the Multimission Archive at the Space Telescope
Science Institute (MAST).  Support for MAST for non-HST data is
provided by the NASA Office of Space Science via grant NNX09AF08G and
by other grants and contracts.

{\it Facilities:} {Swift (UVOT)}, {WIYN:0.9m (S2KB)}, {WIYN
(Sparsepak)}, {Hale (DBSP)}, GBT, Parkes

\end{document}